\documentclass[12pt]{iopart}
\usepackage{iopams}
\usepackage{graphicx}
\usepackage{epsfig}
\usepackage{psfrag}
\usepackage{amssymb}

\def\ra{\rangle}

\def\up{{\mbox{$\uparrow$}}}
\def\down{{\mbox{$\downarrow$}}}

\begin{document}

\title{Preparing multi-partite entanglement of photons and matter qubits}

\author{Pieter Kok\footnote[2]{Electronic address: pieter.kok@hp.com},
Sean D. Barrett\footnote[3]{Electronic address: sean.barrett@hp.com} and
Timothy P. Spiller}

\address{Quantum Information Processing Group, Hewlett-Packard Laboratories, Filton Road, Stoke Gifford, Bristol BS34 8QZ, United Kingdom}

\begin{abstract}
We show how to make event-ready multi-partite entanglement between
 qubits which may be encoded on photons or matter systems.
 Entangled states of matter systems, which can also act as single
 photon sources, can be generated using the entangling operation
 presented in quant-ph/0408040. We show how to entangle such
 sources with photon qubits, which may be encoded in the dual rail,
 polarization or time-bin degrees of freedom. We subsequently
 demonstrate how projective measurements of the matter qubits can
 be used to create entangled states of the photons alone.
 The state of the matter qubits is inherited by
 the generated photons. Since the entangling operation can
 be used to generate cluster states of matter qubits for quantum computing,
 our procedure enables us to create any (entangled) photonic quantum state that can
 be written as the outcome of a quantum computer.
\end{abstract}




\section{Introduction}
\noindent Single-photon sources are a very important resource in
optical quantum communication and computation, and are currently
at the forefront of the experimental effort in optical quantum
information processing
\cite{santori02,beveratos02,yuan02,moreau01,greulich01,michler00,foden00,brunel98,gheri98,Stace2003,Law1997}.
In particular, sources that generate entangled multi-photon states
are very useful, for example in cryptography \cite{ekert91},
linear optical quantum computing \cite{knill01}, and
Heisenberg-limited metrology and quantum lithography \cite{kok04}.
In addition, entanglement between qubits of a different physical
nature (e.g., light and matter) is crucial for distributed quantum
computing.

Here, we consider three alternative representations for encoding
qubits in photonic states. Qubits can be encoded in the
polarization, dual-rail, and time-bin degree of freedom. The first
two are equivalent in that linear optical elements
(polarization beam-splitters, polarization rotators) transform
between the two deterministically.
Here the qubit degree of freedom is the population of two distinct
modes, either two orthogonal polarization modes or two spatial modes,
so a maximally entangled two-qubit state is written as $|H,V\rangle +
|V,H\rangle$ (polarization) or $|0,1;1,0\rangle +
|1,0;0,1\rangle$ (dual-rail). By contrast, the time-bin variable
distinguishes between the arrival times (in one of two bins for a qubit)
of the photon in the
detector. This degree of freedom is particularly useful when the
polarization is subject to decoherence (e.g., in long-distance
quantum communication through optical fibers) \cite{Thew2002}.

In this paper, we present a general method for generating
entangled multi-photon states, by first entangling the single
photon \emph{sources}. The sources we consider have an internal
\emph{matter qubit} degree of freedom (encoded, for example, in
the spin degree of freedom of an electron), and can be entangled
using the double-heralding entangling operation introduced by
Barrett and Kok \cite{barrett04}. This entangling operation can
used to generate cluster states (of the matter qubits) for quantum
computing, and hence can be used to create arbitrary multi-qubit
entangled states \cite{briegel2001,rausschendorf01}. We describe
how these entangled sources can generate photon states that
inherit the entanglement properties of the matter qubits.
Consequently, our method is capable of generating any photonic
quantum state that can be written as the outcome of a quantum
computer. We also show how to create `hetero-entanglement' between
matter qubits and photons. We note that related schemes for
entangling single photon sources (comprising a double-$\Lambda$
energy level configuration), and mapping the entanglement onto
polarization-encoded photonic qubits, have also recently been
proposed \cite{Lim2004a,Lim2004b}. The scheme presented here uses
an alternative energy level scheme, and is also naturally suited
to creating time-bin encoded states.

\section{The single-photon source}

\begin{figure}[t]
  \begin{center}
  \begin{psfrags}
     \psfrag{a}{$|\up\ra$}
     \psfrag{b}{$|\down\ra$}
     \psfrag{c}{$|e\ra$}
     \psfrag{p}[r]{$\pi$-pulse}
     \psfrag{g}{$g$}
       \epsfig{file=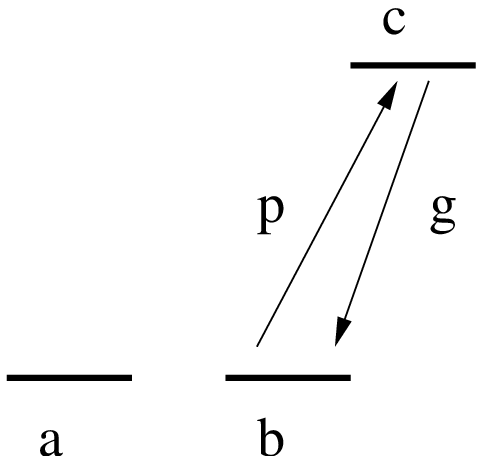, height=5.0cm}
  \end{psfrags}
  \end{center}
  \caption{The qubit system $\{ |\up\ra,|\down\ra\}$ with the excited
  state $|e\ra$. The $\pi$-pulse affects only the transition
  $|\down\ra \to |e\ra$, and the emission of a photon into the cavity
  mode brings the excited state back to the qubit state
  $|\down\ra$. The atom-cavity coupling is given by $g$. The cavity is
  sufficiently leaky in order to release the photon from the cavity
  into the freely propagating mode in the shortest possible time.}
  \label{fig1}
\end{figure}

\noindent The matter system that we consider here is illustrated
in Fig.~\ref{fig1}. The system consists of an atom-like system
comprising three energy levels, with two low-lying levels denoted by
$|\up\rangle$ and $|\down\rangle$ that are (near) degenerate, and
one excited state $|e\rangle$, which is separated from the
low-lying levels by an optical transition. We assume that the
optical transition only couples the levels $|\down\rangle
\leftrightarrow |e\rangle$, and that the transition $|\up\rangle
\leftrightarrow |e\rangle$ is forbidden, e.g. due to a selection
rule. The low lying states $|\up\rangle$ and $|\down\rangle$ may
be thought of as a qubit degree of freedom. We assume that
arbitrary unitary operations can be performed on this qubit, and
furthermore, that one can perform single shot measurements of the
qubit in the computational basis, $\{|\up\rangle,|\down\rangle\}$.

The existence of the optical transition means that the matter
system can also act as an on-demand single photon source. The
three level system is placed in an optical cavity, such that the
cavity mode couples to the $|\down\rangle \leftrightarrow
|e\rangle$ transition. One of the cavity mirrors is assumed to be
leaky, with leakage rate $\kappa$. Applying an optical $\pi$-pulse
(e.g. using a classical laser field) to the system, tuned to the
optical transition, induces the transformation $|\down\ra
\rightarrow |e\ra$ and $|\up\ra \rightarrow |\up\ra$. Provided
that spontaneous emission into modes other than the cavity mode
can be neglected, the matter qubit-cavity system will emit a
single photon into the desired mode, with an emission rate
approximately given by $\Gamma_{\textrm{slow}} =
\kappa-\sqrt{\kappa^2-g^2}$, where $g$ is the Jaynes-Cummings
coupling between the $|\down\rangle \leftrightarrow |e\rangle$
transition and the cavity mode \cite{barrett04}. We define a time
$t_{\textrm{wait}}$ which is the length of time that one should
wait for the observation of a photon in a photo-detector after the
optical $\pi$-pulse 
has been applied. Provided $t_{\textrm{wait}} \gg
\Gamma_{\textrm{slow}}^{-1}$, and the initial state of the source
was $|\down\ra$, a single photon wavepacket will be emitted into
the desired mode with probability very close to one. Note that the
emission of a photon from this system is conditional on the
initial state of the qubit degree of freedom. In some physical
implementations, it may be advantageous to use a fourth auxiliary
level, such that the $|\down\rangle \leftrightarrow |e\rangle$
transition is replaced by a STIRAP process in a $\Lambda$
configuration of three levels \cite{Law1997}.

\section{Entangling matter qubits}

Matter qubits (with a computational basis
$\{|\up\rangle,|\down\rangle\}$) can be prepared in entangled
states using the {\em double-heralding} technique described in
\cite{barrett04}. We briefly review this technique here. We
entangle two qubits by first preparing two atoms in separate
cavities in the separable state $(|\up\rangle +
|\down\rangle)(|\up\rangle + |\down\rangle)$. Subsequently, we
apply an optical $\pi$-pulse to each atom, and wait for a time
$t_{\textrm{wait}}$. This yields the total state
\begin{equation}
 |\up\up\rangle |0,0\rangle + |\up\down\rangle |0,1\rangle +
  |\down\up\rangle |1,0\rangle + |\down\down\rangle |1,1\rangle \; ,
\end{equation}
where $|0\rangle$ and $|1\rangle$ respectively denote the vacuum and a
single photon wavepacket in the freely propagating optical mode leaving the
cavity. When these two modes interact in a 50:50 beam splitter, the
total state becomes
\begin{equation}
 |\up\up\rangle |0,0\rangle + \frac{1}{\sqrt{2}}
  \Bigl[(|\up\down\rangle + |\down\up\rangle) |0,1\rangle  +
  (|\up\down\rangle - |\down\up\rangle) |1,0\rangle +
  |\down\down\rangle (|2,0\rangle - |0,2\rangle) \Bigr] \; . \nonumber
\end{equation}
Note that the beam splitter must be placed in such a way that the
spatio-temporal photon modes overlap at the beam splitter, in
order to erase the `which path' information. Detecting both the
outgoing modes of the beam splitter, each with a realistic
detector (i.e. a detector with finite efficiency, and which cannot
discriminate between optical states with one or more photons),
gives the following state of the qubits (given just a single
detector click):
\begin{equation}\label{intermediate}
 \rho = f(\eta) |\Psi_{\pm}\rangle\langle\Psi_{\pm}| + (1-f(\eta))
 |\down\down\rangle\langle\down\down|\; ,
\end{equation}
where $|\Psi_{\pm}\rangle = (|\up\down\rangle \pm
|\down\up\rangle)/\sqrt{2}$ and $f(\eta)\leq 1$ is a function of
the combined collection and detection efficiency, $\eta$. The
relative phase in $|\Psi_{\pm}\rangle$ is determined by the
detection signature, (``click'',``no click'') or (``no
click'',``click'') for the two detectors.

The state in Eq.~(\ref{intermediate}) is an incoherent mixture of
a maximally entangled state and the separable state
$|\down\down\rangle\langle\down\down|$. However, we can remove
this separable part by first applying a bit flip operation
$|\up\rangle \leftrightarrow |\down\rangle$ to both matter qubits.
We subsequently apply
a second $\pi$-pulse to each matter system. The separable part
cannot generate photons. Thus, conditional on observing another single
detector click, the final two qubit state
\begin{equation}
 |\Psi\rangle = \frac{1}{\sqrt{2}}(|\up\down\rangle \pm
 |\down\up\rangle)\; \label{eq:PerfectEntanglement}
\end{equation}
is obtained. The total success probability of this procedure is
$\eta^2/2$.  Note that we can also make {\em any} pure two-qubit
state by using the initial states $\mu_j |\up\ra + \nu_j
|\down\ra$ (rather than $(|\up\rangle + |\down\rangle)$),
performing the double heralding steps described above, and
subsequently performing single-qubit operations. Furthermore,
using this entangling operation, combined with single qubit
unitaries and measurements, we can efficiently produce cluster
states of \emph{many} matter qubits \cite{barrett04}. This is an
extremely useful result, since it is known that many other interesting
multi-qubit entangled states, such as GHZ states, can be generated
by performing single qubit measurements on cluster states
\cite{briegel2001}. But most importantly, cluster states together with
single qubit unitary operations and measurements can be used to
implement any quantum algorithm \cite{rausschendorf01}. We can
therefore efficiently produce any state that may be the output of 
a quantum computer.

It is worth discussing here the effect of experimental
imperfections on the state given in Eq.
(\ref{eq:PerfectEntanglement}), which of course represents an
idealization of the state of the two qubit system after the {\em
double-heralding} operation. In real systems, physical
imperfections can reduce the fidelity of the state. Perhaps the
most important such imperfection is that of photon loss, for
example by emission into unwanted modes, absorption in the optical
elements, or detector inefficiency. It turns out that such losses
do not affect the fidelity of our entangled output states, but
merely reduce the success probability of the protocol
\cite{barrett04}. Mismatch in the cavity parameters $g$ and
$\kappa$ between different cavities can reduce the fidelity.
However this scheme is reasonably robust to such losses: a
mismatch of a few percent leads to a reduction in fidelity of less
than $10^{-3}$ \cite{barrett04}. Decoherence of the qubit degrees
of freedom can also reduce fidelity. However this can be mitigated
by choosing a system whose intrinsic decoherence time is long
compared to $t_{\textrm{wait}}$; examples of such systems were
given in \cite{barrett04}.

\section{Entanglement between matter qubits and photons}

\noindent We can now create entangled states between the matter
qubits and photons by applying more $\pi$-pulses (and bit flips).
This type of `hetero-entanglement' (entanglement between systems
of a different physical class) can be very useful, for example in quantum
key distribution via the Ekert protocol \cite{ekert91}. In that
case, it is sufficient for Alice and Bob to share {\em any}
maximally entangled two-qubit state, such as $|\up\rangle
|H\rangle + |\down\rangle |V\rangle$ (where $H$ and $V$ denote
polarized photons). Alice can measure her matter qubit in a spin
basis of her choice, while Bob uses photo-detection in a
polarization basis that he chooses.

\begin{figure}[t]
  \begin{center}
  \begin{psfrags}
     \psfrag{Alice}{\underline{Alice}}
     \psfrag{Bob}{\underline{Bob}}
     \psfrag{pbs}{\sc pbs}
     \psfrag{H}{``$H$''}
     \psfrag{V}{``$V$''}
     \psfrag{a}{a)}
     \psfrag{b}{b)}
     \psfrag{t}{``$t_{\mathrm{arrival}}$''}
       \epsfig{file=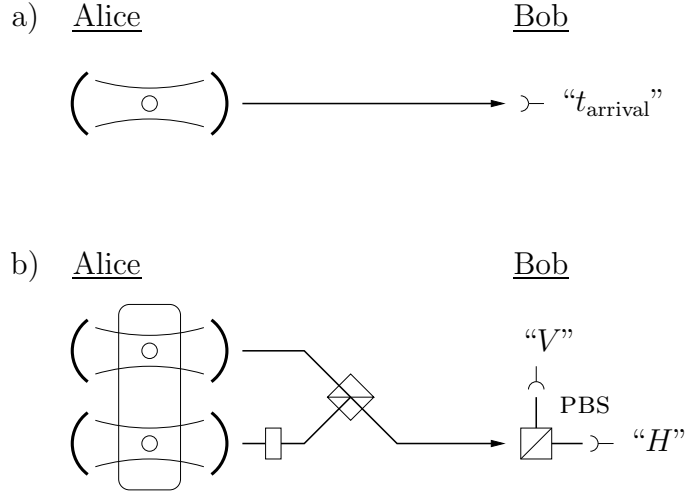,width=8.5cm}
  \end{psfrags}
  \end{center}
  \caption{Creating entanglement between the qubit in the cavity and a
  freely propagating photon: a) for time-bin--qubit entanglement we
  apply a $\pi$-pulse, wait $t_{\textrm{wait}}$ (the early bin),
  apply a bit flip and a second $\pi$-pulse, and wait $t_{\textrm{wait}}$
  (the late bin);
  b) for polarization--qubit entanglement we apply a single $\pi$-pulse.
  In b) a polarization rotation (bit flip) is applied to one mode before
  they are combined on a polarization beam-splitter (PBS) for transmission.
  Bob uses a PBS at his end to separate the polarizations for detection.}
  \label{fig2}
\end{figure}

Alice and Bob can share hetero-entanglement using time-bin
photons, polarized photons, or dual rail photons. To generate
time-bin entanglement, let Alice hold a cavity with a matter qubit
in the state $|\up\rangle + |\down\rangle$. After applying an
optical $\pi$-pulse, waiting for a time $t_{\textrm{wait}}$,
applying a bit flip, applying another optical $\pi$-pulse, and
waiting for a second time window $t_{\textrm{wait}}$, the total
state will be given by $|\up\rangle |E\rangle + |\down\rangle
|L\rangle$, where $|E\ra$ and $|L\ra$ denote photon wavepackets
localized in the `early' and `late' spatio-temporal modes,
respectively. The optical mode that supports the photon is
detected by Bob (see Fig.~\ref{fig2}a). Note that, when qubits are
represented by such time-bin photons, Alice and Bob must share
some classical timing reference information, such that both
parties agree on the definition of `early' and `late' time bins.
This information could be provided by, for example, a shared
classical laser pulse or electronic signal.

Alternatively, we can entangle the qubit with the polarization
degree of freedom of a photon. This variant requires two matter
qubits: Alice first prepares the two matter qubits in the
maximally entangled state $|\down\up\rangle + |\up\down\rangle$.
This is formally equivalent to a {\em new} qubit $|\tilde{0}\ra +
|\tilde{1}\ra$, with $|\tilde{0}\ra \equiv |\down\up\rangle$ and
$|\tilde{1}\ra \equiv |\up\down\rangle$. The next step in
generating hetero-entanglement is then to apply a $\pi$-pulse to
the two matter qubits, and waiting for a time $t_{\textrm{wait}}$
for a photon to be emitted. By construction, a photon will be
emitted by one and only one of the matter systems. At this point
we have generated hetero-entanglement between the composite qubit
system and a dual-rail photon. Assuming that the outgoing photon
has a definite polarization (e.g., horizontal), one of the modes
undergoes a polarization rotation, and the two modes are combined
with a polarization beam splitter (see Fig.~\ref{fig2}b). The
total system will be in the state $|\tilde{0}\rangle |H\rangle +
|\tilde{1}\rangle |V\rangle$.

Note that single qubit operations on \emph{physical} qubits are
not sufficient to perform arbitrary operations in the
\emph{encoded} basis $\left\{|\tilde{0}\ra,|\tilde{1}\ra\right\}$,
as may be required for some applications. However, it is possible
to remove one of the physical qubits by performing a measurement
in the $\left\{ |+\ra,|-\ra \right\}$ basis, where $|\pm\ra =
|\up\rangle \pm |\down\rangle$. Such a measurement can be
implemented by first performing a Hadamard operation on the
relevant physical qubit, and subsequently performing a measurement
in the computational basis. The resulting state is of the form
$|\down\rangle |H\rangle \pm |\up\rangle |V\rangle$. Note that this
state contains only one matter qubit, which can be manipulated
using single qubit operations. The relative phase between the
terms is determined by the outcome of the physical qubit
measurement. This conditional phase can be corrected using a
single qubit operation on either the matter qubit or the photon,
provided sufficiently fast classical switching is available.
Alternatively, it is sufficient in many applications just to keep
a (classical) record of the relative phase, and take it into
account when interpreting the results of individual measurements
of the entangled qubits.

\section{Entangled states of light}

The entangled states of matter qubits can also be used to make
entangled states of light. Suppose we want to make time-bin
entanglement of the form $|E,L\rangle + |L,E\rangle$. We start
with the two-qubit state $|\down\up\rangle + |\up\down\rangle$ and
apply a $\pi$-pulse to the two cavities. After a time $t_{\textrm{wait}}$
the total state is then
\begin{equation}\nonumber
 |\down\up\rangle |E,0\rangle + |\up\down\rangle |0,E\rangle\; .
\end{equation}
That is, the early photon wavepacket 
$|E\rangle$ has left the cavities. We then
perform a bit-flip on the qubits, after which we apply the second
$\pi$-pulse to both cavities. After a further time $t_{\textrm{wait}}$,
this yields the highly entangled state
\begin{equation}\nonumber
 |\up\down\rangle |E,L\rangle + |\down\up\rangle |L,E\rangle\; .
\end{equation}
This may be thought of as four-party hetero-entanglement.

\begin{figure}[t]
  \begin{center}
  \begin{psfrags}
     \psfrag{Alice}{\underline{Alice}}
     \psfrag{Bob}{\underline{Bob}}
     \psfrag{pbs}{\sc pbs}
     \psfrag{H}{``$H$''}
     \psfrag{V}{``$V$''}
     \psfrag{a}{a)}
     \psfrag{b}{b)}
       \epsfig{file=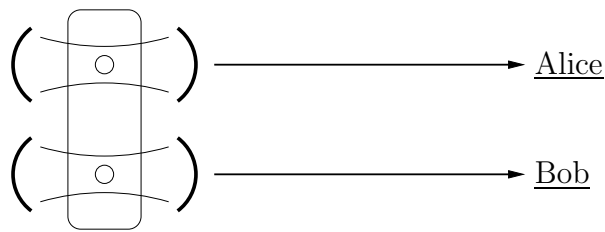, width=8cm}
  \end{psfrags}
  \end{center}
  \caption{Creating two time-bin entangled photons. We start with the
  two matter qubits in the entangled state $|\up\down\rangle +
  |\down\up\rangle$. The first
  $\pi$-pulse in the cavity system yields the early photon. After a
  bit flip on both qubits, the second $\pi$-pulse yields the late
  photon.}
  \label{fig3}
\end{figure}

When we want two-photon entanglement, we have to {\em disentangle} the
photons from the matter qubits. To this end, we apply a Hadamard
operation to both matter qubits. The total state then becomes
\begin{eqnarray}
 |\Psi\rangle &=& \frac{1}{2\sqrt2}|\up\up\rangle (|E,L\rangle + |L,E\rangle) \cr
  && +  \frac{1}{2\sqrt2}|\up\down\rangle (-|E,L\rangle + |L,E\rangle) \cr
  && +  \frac{1}{2\sqrt2}|\down\up\rangle (|E,L\rangle - |L,E\rangle) \cr
  && -  \frac{1}{2\sqrt2}|\down\down\rangle (|E,L\rangle + |L,E\rangle)\; .
\end{eqnarray}
A measurement of the matter qubits in the computational basis will
then reveal which particular form of entanglement we have
prepared. As in the case of hetero-entanglement between matter
qubits and polarization qubits, when the measurement outcome is
$\up\down$ or $\down\up$, the relative minus sign can be directly
corrected using a fast-switching phase shift on one of the qubits,
or by taking account of the relative phase when interpreting
subsequent measurement outcomes. When the output of the two
cavities is sent to Alice and Bob, respectively, they will share
maximal two-photon time-bin entanglement (see Fig.~\ref{fig3}).

\begin{figure}[t]
  \begin{center}
  \begin{psfrags}
     \psfrag{Alice}{\underline{Alice}}
     \psfrag{Bob}{\underline{Bob}}
       \epsfig{file=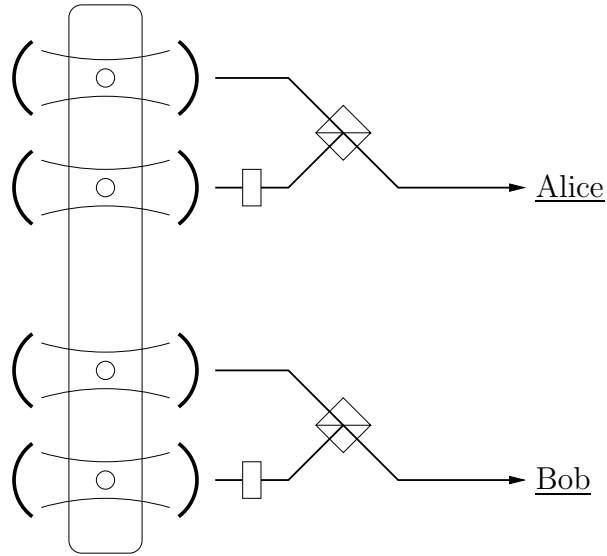, width=8cm}
  \end{psfrags}
  \end{center}
  \caption{Creating polarization entanglement using four entangled
  matter qubits. These four qubits are prepared in the GHZ state
  $|\up\down\down\up\rangle + |\down\up\up\down\rangle$. This will
  create dual-rail entangled photons after a $\pi$-pulse to the
  four-qubit system. As in Fig.~2b), one photon qubit mode undergoes a
  polarization rotation before the modes are combined at a PBS for
  transmission.}
  \label{fig4}
\end{figure}

In order to make maximal dual rail and polarization entanglement,
we need four matter qubits in four separate leaky cavities (see
Fig.~\ref{fig4}). A $\pi$-pulse applied to each of these systems
will yield the following transformation, after a time $t_{\textrm{wait}}$:
\begin{eqnarray}
 |\tilde{0}\rangle \otimes |0,0\rangle &\rightarrow& |\tilde{0}\rangle
  \otimes |1,0\rangle \cr
 |\tilde{1}\rangle \otimes |0,0\rangle &\rightarrow& |\tilde{1}\rangle
  \otimes |0,1\rangle\; .
\end{eqnarray}
where we have used the composite qubit systems $|\tilde{0}\rangle$ and
$|\tilde{1}\rangle$. The states $|0,1\rangle$ and $|1,0\rangle$ denote
a single photon with a dual-rail degree of freedom.

The four matter qubits must first be prepared in the state
$|\tilde{0}, \tilde{1}\rangle + |\tilde{1},\tilde{0}\rangle$. This
state corresponds to a GHZ state of four matter qubits which, as
noted above, may be generated by first preparing a larger cluster
state and subsequently performing single qubit measurements.
Subsequently, applying a $\pi$-pulse to each matter qubit
generates a highly entangled state
\begin{equation}\nonumber
 |\tilde{0},\tilde{1}\rangle |1,0;0,1\rangle + |\tilde{1},\tilde{0}\rangle
  |0,1;1,0\rangle\; .
\end{equation}
Again, we have to transform each of the four matter qubits with a
Hadamard gate, and perform a measurement of the four matter qubits in
the computational basis. Up to correctable relative phases, we have
obtained the required entangled state.

There is a duality between the dual-rail representation and the
time-bin representation. In the dual-rail (or polarization)
representation, we need $2N$ matter qubits in a suitable state to
create an $N$-photon entangled state $|\psi\rangle$. On the other
hand, in the time-bin representation we need only $N$ matter qubits,
but we need to apply the $\pi$-pulse twice (with an intermediary bit
flip on all the qubits). In their respective computational bases,
these representations are thus completely equivalent.

\section{Multi-photon entanglement}

\noindent
The definition of $|\tilde{0}\rangle$ and $|\tilde{1}\rangle$ is
particularly fruitful, in that it immediately suggests a
generalization of the schemes presented here. We consider the
generation of polarized photonic states. Let $|{\mathcal{P}}_k\rangle$ be
a string of $N$ photonic qubit states in the computational basis (for
example, when $N=2$, the four strings $|{\mathcal{P}}_k\rangle$ for polarized
photons are $|H,H\rangle$, $|H,V\rangle$, $|V,H\rangle$, and
$|V,V\rangle$). Any pure (entangled) state can then be written as
\begin{equation}\nonumber
 |\psi\rangle = \sum_{k=1}^{2^N} \alpha_k |{\mathcal{P}}_k\rangle
\end{equation}
In order to create such a state, we need to create a $2N$-qubit state
\begin{equation}\label{multi}
 |\Psi\rangle = \sum_{k=1}^{2^N} \alpha_k |{\mathcal{S}}_k\rangle
  \otimes |0\rangle_P\, ,
\end{equation}
where ${\mathcal{S}}_k$ is a string of $N$ bipartite qubits
$|\tilde{0}\rangle$ and $|\tilde{1}\rangle$ and $|0\rangle_P$ the
vacuum of $N$ spatial modes. If such a state can be obtained with
a quantum computer, we can use the technique in \cite{barrett04}
to create the appropriate cluster state and do the computation.
The final state is then given by Eq.~(\ref{multi}).

A $\pi$-pulse into all $2N$ cavities will then create the photons, which
are still heavily entangled with the qubits:
\begin{equation}\nonumber
 |\Psi\rangle \rightarrow \sum_{k=1}^{2^N} \alpha_k |{\mathcal{S}}_k\rangle
  \otimes |{\mathcal{P}}_k\rangle_P\, .
\end{equation}
Applying local Hadamard operations to all matter qubits and
measuring them in the computational basis then yields the desired
state up to local single-photon transformations.
\begin{equation}\nonumber
 |\psi_{\ell}\rangle_P = \mathrm{Tr}_{\mathcal{S}} \left[ H^{\otimes 2N}
  |\Psi\rangle\langle\Psi|H^{\otimes 2N} \hat{D}_{\ell}\right] ,
\end{equation}
where $D_{\ell}$ is the detector signature of all $2N$ matter
qubits, and $H$ is the Hadamard transform of a single matter
qubit. There are $4^N$ possible detector outcomes (indexed by
$\ell$), and the output state $|\psi_{\ell}\rangle_P$ can be
transformed into $|\psi\rangle$ with  phase shifters and
polarization rotations. The procedures described above for
generating dual rail or time-bin entanglement can similarly be
generalized to the multi-qubit case.

It is worth noting briefly the effect that photon loss has on the
scalability of the scheme. Photon loss can, ultimately, be
detected, since in many applications, all the photons can be
detected at the end of the experiment. Thus photon loss in itself
need not degrade the fidelity of the resulting entangled states.
However, the total success probability (i.e. the probability of
detecting all $N$ photons) scales exponentially with $N$. This
will ultimately place an upper limit on $N$ in a practical
experiment. Nevertheless, even relatively small numbers of
entangled photonic qubits can be a useful resource in quantum
information processing tasks such as cryptography \cite{ekert91} and
Heisenberg-limited metrology \cite{kok04}.
Furthermore, they can be used as a resource in linear optical
quantum computing schemes \cite{knill01}, which are in
principle scalable.

\section{Conclusion}

In conclusion, we have shown how the double-heralded entangling
operation introduced by Barrett and Kok \cite{barrett04} can be
used to create multi-partite photonic entanglement. In particular,
we can create any (entangled) state of photonic qubits that can be
written as the outcome of a universal quantum computer. This works in the
polarization and dual rail basis, as well as in the time-bin
degree of freedom. In addition, we can generate
hetero-entanglement between (many) qubits of a different physical
nature. This is crucial for distributed quantum computing, where
local clusters of matter qubits can be connected via optical
(flying) qubits.

\vskip 2 truecm \noindent {\em Acknowledgments}: PK and SDB have
contributed equally to this work. We thank Rob Thew for bringing
the virtues of time-bin entanglement to our attention, and Yuan Liang
Lim and Almut Beige for fruitful discussions. The authors are
supported by the EU Nanomagiq and Ramboq projects.

\end{document}